\documentclass[fleqn,usenatbib,twocolumn]{mnras}
\usepackage{newtxtext,newtxmath}
\usepackage[T1]{fontenc}
\usepackage{ae,aecompl}
\usepackage{graphicx}
\usepackage{amsmath}	
\usepackage{amssymb}	
\usepackage{hyperref}
\begin{document}

\title[Spikey in the radio]{A self-lensing supermassive binary black hole at radio frequencies: the story of Spikey continues}

\author[Kun et al.]{
Emma Kun$^{1}$\thanks{Email: kun.emma@csfk.mta.hu}, 
S\'{a}ndor Frey$^{1,2}$
and
Krisztina \'{E}. Gab\'{a}nyi$^{3,4,1}$
\vspace{2mm}
\\ 
$^{1}$ Konkoly Observatory, Research Centre for Astronomy and Earth Sciences, Konkoly Thege Mikl\'{o}s \'{u}t 15-17, H-1121 Budapest, Hungary \\
$^{2}$ Institute of Physics, ELTE E\"otv\"os Lor\'and University, P\'azm\'any P\'eter s\'et\'any 1/A, H-1117 Budapest, Hungary \\
$^{3}$ Department of Astronomy, E\"{o}tv\"{o}s Lor\'{a}nd University, P\'{a}zm\'{a}ny P\'{e}ter s\'{e}t\'{a}ny 1/A, H-1117 Budapest, Hungary \\
$^{4}$ MTA-ELTE Extragalactic Astrophysics Research Group, P\'{a}zm\'{a}ny P\'{e}ter s\'{e}t\'{a}ny 1/A, H-1117 Budapest, Hungary 
}

\date{}

\maketitle

\begin{abstract}
The quasar J1918+4937 was recently suggested to harbour a milliparsec-separation binary supermassive black hole (SMBH), based upon modeling the narrow spike in its high-cadence {\em Kepler} optical light curve. Known binary SMBHs are extremely rare, and the tight constraints on the physical and geometric parameters of this object are unique. The high-resolution radio images  of J1918+4937 obtained with very long baseline interferometry (VLBI) indicate a rich one-sided jet structure extending to 80 milliarcseconds. Here we analyse simultaneously-made sensitive 1.7- and 5-GHz archive VLBI images as well as snapshot 8.4/8.7-GHz VLBI images of J1918+4937, and show that the appearance of the wiggled jet is consistent with the binary scenario. We develop a jet structural model that handles eccentric orbits. By applying this model to the measured VLBI component positions, we constrain the inclination of the radio jet, as well as the spin angle of the jet emitter SMBH. We find the jet morphological model is consistent with the optical and radio data, and that the secondary SMBH is most likely the jetted one in the system. Furthermore, the decade-long 15-GHz radio flux density monitoring data available for J1918+4937 are compatible with a gradual overall decrease in the the total flux density caused by a slow secular change of the jet inclination due to the spin--orbit precession. J1918+4937 could be an efficient high-energy neutrino source if the horizon of the secondary SMBH is rapidly rotating.
\end{abstract}


\section{Introduction}
\label{intro}

Recently \citet{hu20} interpreted a narrow spike in the densely-sampled {\em Kepler} optical light curve of the quasar J1918+4937 \citep[also known as KIC~11606854, dubbed as Spikey by][]{hu20} as a result of gravitational self-lensing in a supermassive black hole binary (SMBHB) system. The quasar has a spectroscopic redshift of $z_\mathrm{sp}=0.926$ \citep{healey08}. In this scenario, the orbital plane of the binary lies sufficiently close to the line of sight so that when one of the companions -- the black hole with the larger mass -- passes in front of the other, the optical emission of the latter active galactic nucleus (AGN) is significantly enhanced. Taking two relativistic effects, the binary self-lensing and the orbital Doppler boosting into account, \citet{hu20} modeled the {\em Kepler} light curve containing the spike. They found that the system is composed of two black holes (BHs), with masses of $2.5 \times 10^7$~M$_\odot$ and $5.0 \times 10^6$~M$_\odot$. The eccentric orbit ($e \approx 0.52$) has a period of $T=418$~d in the rest frame of the object. From our point of view, the orbital plane is seen almost edge-on, within an angle of $\sim 8\degr$.

Studying binary AGNs is an active field of both observational and theoretical astrophysics, due to its connection to cosmological structure formation, galaxy evolution, and most recently gravitational waves. Observations of such objects are very challenging \citep[for a review see e.g.][]{Komossa2016}, and securely confirmed cases are extremely rare \citep{derosa19}. Spikey stands out from the very few SMBHB candidates because the binary self-lensing model \citep{hu20} constrains the orbital parameters, the geometry, and the masses of the companions very accurately. The model also provides a testable prediction that the next flaring will occur in 2020. 

Apart from being a moderately bright X-ray AGN \citep{hu20}, J1918+4937 is also a prominent radio-loud quasar. Variations in its $\sim100$-mJy level flux density at 15~GHz are being monitored at the Owens Valley Radio Observatory \citep[OVRO,][]{richards11}. The source is known to have a compact radio jet structure at milliarcsec (mas) angular scales, as revealed by very long baseline interferometry (VLBI) imaging observations \citep[e.g.][]{kovalev07}. Detecting binary AGNs separated by a small fraction of a pc is practically impossible with direct imaging observations, even with the high resolution offered by VLBI and in the most nearby universe \citep[e.g.][]{an2018}. But radio interferometric observations could help in another way, by detecting a discernible effect of a binary companion on the appearance of the relativistic jet produced by the other AGN in the system, because the orbital motion of the jetted AGN may result in a helical shape of the jet \citep[for a recent review, see][and references therein]{derosa19}.

A growing number of studies propose radio-loud AGNs as strong candidates for efficient high-energy (HE) neutrino emitter objects, especially the blazars \citep[e.g.][]{Kadler2016,ICTXS2018a,Garrappa2019,Giommi2020}, whose jets point close to the line of sight of the observer. The underlying physical mechanisms involve light--matter and/or matter--matter interactions in a relativistically moving plasma. 
\citet{Kun2017,Kun2019} proposed a model in which the radio and neutrino observations were put into a common physical picture involving the spin-flip of a SMBH in a merging binary. Recently the $\gamma$-ray flaring blazar TXS~0506+056, an efficient particle accelerator, turned out to be the source of several IceCube neutrinos \citep{ICTXS2018a,ICTXS2018b}. Studies indicate that neutrino emission might be due to a recent merger activity \citep{Britzen2019,Kun2019}. Although there is no indication yet of an observed neutrino event near its position, Spikey is a VLBI source directing its jet close to our line of sight, and also a SMBH binary candidate, making this AGN an object of great interest as a potential neutrino source.

In this paper, we investigate whether the available radio data are consistent with the behaviour of the object, the gravitational self-lensing model, and in particular the parameters derived for Spikey by \citet{hu20}. Based on archival data from 2008, we present sensitive and detailed VLBI images of J1918+4937 obtained at 1.7 and 5~GHz for the first time, and show the OVRO flux density curve (Sect.~\ref{radio}). By modeling the source brightness distribution at $\sim1-10$~mas scale, we derive parameters describing the relativistic jet, estimate the apparent speed of the jet based on snapshot VLBI observations conducted at the 8.4/8.7 GHz frequency band, and put forward a scenario where a jet is launched from one of the accreting BH components of the system in Sect.~\ref{jetmodel}. Here we also investigate the case whether the OVRO flux density curve is compatible with the binary model. We discuss our findings based on the OVRO single-dish and VLBI radio observations in Sect.~\ref{discussion}. We also discuss whether Spikey could be an efficient high-energy neutrino emitter in the near future based on the behaviour of its VLBI jet and the proposed SMBH merger scenario. Finally we conclude the paper with a summary in Sect.~\ref{summary}.

We assume a flat $\Lambda$CDM cosmological model with $H_{\rm 0}$=70~km~s$^{-1}$~Mpc$^{-1}$, $\Omega_{\rm m}=0.3$, and $\Omega_{\Lambda}=0.7$ in this paper. In this model, an object at $z_\mathrm{sp}=0.926$ has a luminosity distance of $D_{\rm L} \approx 6$~Gpc, and 1~mas angular size corresponds to $7.853$~pc projected linear size \citep{wright06}.

\section{Radio observations}
\label{radio}

\subsection{VLBI imaging observations and data reduction}
\label{vlbi}

\citet{kharb10} studied the Seyfert galaxy NGC~6764 with high-resolution VLBI imaging at 1.7 and 5~GHz. The observations were conducted with the U.S. Very Long Baseline Array (VLBA) in phase-referencing mode \citep{beasley95} where J1918+4937 (1917+495) was selected as the nearby compact calibrator source within $2\degr$ separation from the target. Nine 25-m diameter antennas of the VLBA (Brewster, Fort Davis, Hancock, Kitt Peak, Los Alamos, Mauna Kea, North Liberty, Owens Valley, and Pie Town) participated in the experiment BK154, performed on 2008 November 13-14 with a total duration of about 14~h. The observations at both frequencies were made with two 8-MHz wide intermediate frequency channels (IFs) in both left and right circular polarizations. The total bandwidth was therefore 32~MHz. 

\citet{kharb10} scheduled the observations in 5-min switching cycles, with 2~min spent on the calibrator J1918+4937 and 3~min on the weak target NGC~6764, including antenna slewing times. As a valuable byproduct of this phase-referencing experiment, $\sim$1.6 and 1.8~h of VLBA data accumulated on our object of interest, J1918+4937, at 1.7 and 5~GHz, respectively. 

We downloaded the raw VLBA data of the BK154 experiment from the public archive\footnote{\url{https://archive.nrao.edu}} of the U.S. National Radio Astronomy Observatory (NRAO). For the data calibration, we used the NRAO Astronomical Image Processing System \citep[{\sc AIPS},][]{greisen03} in a standard way \citep[e.g.][]{diamond95}. We started with calibrating the ionospheric delays based on total electron content measurements, and corrected for the measured Earth orientation parameters. We applied digital sampler corrections, then used a short 1-min scan on the bright fringe-finder source 1758+388 to solve for instrumental phases and delays. Bandpass correction was performed using the same scan. We used the gain curve and system temperature information from the participating VLBA stations for a-priori amplitude corrections. Finally fringe-fitting was done and the solutions were applied to the data. 

\begin{figure*}
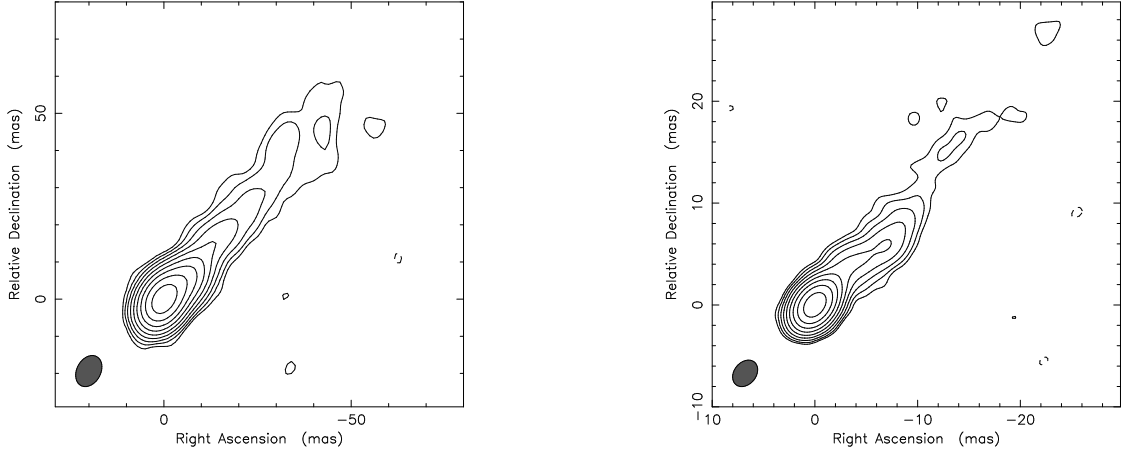

\centering
\includegraphics[bb=65 95 504 720, width=60mm, angle=270, clip=]{1917+495-Lband.ps}
\includegraphics[bb=65 95 504 720, width=60mm, angle=270, clip=]{1917+495-Cband-zoom.ps}
\caption{VLBA images of J1918+4937. {\em Left:} at 1.7~GHz. The peak intensity is 128~mJy\,beam$^{-1}$. The lowest contours are drawn at $\pm 0.3$~mJy\,beam$^{-1}$. The elliptical Gaussian restoring beam is 8.9~mas\,$\times$\,6.4~mas (FWHM) at a position angle $-28\degr$. {\em Right:} at 5~GHz. The peak intensity is 119~mJy\,beam$^{-1}$. The lowest contours are drawn at $\pm 0.28$~mJy\,beam$^{-1}$. The restoring beam is 2.9~mas\,$\times$\,2.1~mas (FWHM) at a position angle $-38\degr$. In both images, the positive contour levels increase by a factor of 2, and the restoring beam size is indicated in the bottom-left corner.}
\label{images}
\end{figure*}

The calibrated visibility data of J1918+4937 were exported to the {\sc Difmap} software \citep{shepherd94}. After the standard hybrid mapping procedure involving several iterations of {\sc clean} decomposition, phase-only self-calibration, and finally phase and amplitude self-calibration, we obtained the naturally-weighted 1.7- and 5-GHz VLBI images of J1918+4937 shown in Fig.~\ref{images}. 
As a finishing step of data reduction, we fitted circular Gaussian brightness distribution model components directly to the self-calibrated visibility data in {\sc Difmap}. This allows us to describe the radio structure with a limited set of parameters that are listed in Table~\ref{modelfit}, where the errors were calculated as in \citet{Kun2014}. These model components will be used for determining the shape of the jet. This way we can also gain information on the Doppler boosting of the relativistic jet.  

\begin{table*}
  \caption[]{Parameters of the circular Gaussian model components fitted to the 1.7 and 5 GHz VLBI visibility data of J1918+4937.}
  \label{modelfit}
\begin{tabular}{ccccc}        
\hline                 
Frequency       & Flux density  & \multicolumn{2}{c}{Relative position}& Diameter     \\
(GHz)           & $F$ (mJy)     & R.A. (mas)        & Dec. (mas)       & FWHM (mas)   \\
\hline                       
1.7    &$104 \pm 6$ & $0.00 \pm 0.34$ & $0.00 \pm 0.41$ & $0.57 \pm 0.01$\\
        &$30 \pm 2$ & $-2.07 \pm 0.36$ & $2.56 \pm 0.42$ & $1.18 \pm 0.01$\\
        &$24 \pm 2$ & $-6.03 \pm 0.39$ & $5.29 \pm 0.45$ & $2.02 \pm 0.01$\\
        &$7 \pm 1$ & $-9.07 \pm 0.38$ & $9.23 \pm 0.44$ & $1.73 \pm 0.01$\\
        &$5 \pm 1$ & $-14.36 \pm 0.46$ & $16.51 \pm 0.51$ & $3.17 \pm 0.08$\\
        &$4 \pm 1$ & $-21.76 \pm 0.97$ & $23.79 \pm 0.99$ & $9.08 \pm 0.15$\\
        &$5 \pm 2$ & $-34.95 \pm 1.71$ & $40.20 \pm 1.73$ & $16.80 \pm 2.73$\\
\hline
5      & $100 \pm 7$ & $0.00 \pm 0.12$ & $0.00 \pm 0.13$ & $0.18 \pm 0.01$\\
        & $32 \pm 4$ & $-0.81 \pm 0.12$ & $1.09 \pm 0.13$ & $0.33 \pm 0.01$\\
        & $6 \pm 2$ & $-2.70 \pm 0.18$ & $3.43 \pm 0.19$ & $1.39 \pm 0.02$\\
        & $6 \pm 1$ & $-4.85 \pm 0.20$ & $4.90 \pm 0.20$ & $1.59 \pm 0.05$\\
        & $4 \pm 1$ & $-7.26 \pm 0.17$ & $6.06 \pm 0.17$ & $1.17 \pm 0.02$\\
        & $3 \pm 1$ & $-8.43 \pm 0.28$ & $8.28 \pm 0.28$ & $2.54 \pm 0.03$\\
        & $3 \pm 1$ & $-13.45 \pm 0.70$ & $16.71 \pm 0.70$ & $6.86 \pm 0.71$\\
\hline
\end{tabular}
\end{table*}

\begin{figure}
\centering
\includegraphics[angle=270,scale=0.5]{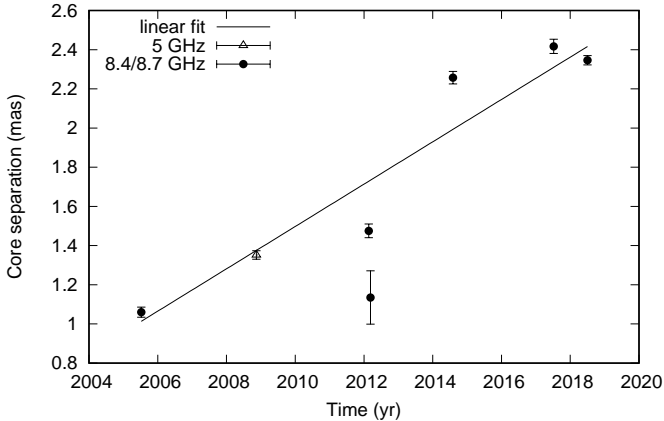}
\caption{Separation of the core and the inner jet component identified at each epoch in J1918+4937 as a function of time, based on 8.4/8.7-GHz (filled circles) and 5-GHz (open triangle) VLBI measurements. The line represents the linear fit for estimating the apparent angular proper motion, $0.11 \pm 0.01$~mas\,yr$^{-1}$.}
\label{fig:propermotion}
\end{figure}

Further VLBI imaging observations of J1918+4937 made at the 8.4/8.7-GHz frequency band are available in the Astrogeo data base\footnote{\url{http://astrogeo.org/cgi-bin/imdb_get_source.csh?source=J1918\%2B4937}} covering a $13$-yr long interval from 2005 to 2018. These are short snapshot VLBA observations of varying quality, typically with scans of few minutes, not suitable for recovering the fine details of the jet structure. We downloaded the calibrated visibility data and performed imaging and model fitting in {\sc Difmap}. At each epoch, we could at least model the core emission and the innermost jet component with circular Gaussian brightness distribution components, allowing us to measure their angular separation. The values are given in Table~\ref{x-modelfit} and plotted as a function of time in Fig.~\ref{fig:propermotion}.
We also included here the 5-GHz data point (Table~\ref{modelfit}) because it was obtained at a close frequency and it helps filling the gap in the time coverage of the 8.4/8.7-GHz measurements. Although the data points have a scatter beyond the formal uncertainties due to the complications with imaging a complex structure from limited observations, the component separation clearly increases with time. The line in Fig.~\ref{fig:propermotion} indicates the linear fit for estimating the apparent angular proper motion, $0.10 \pm 0.01$~mas\,yr$^{-1}$.

\begin{table}
  \centering 
  \caption[]{Separation of the core and the innermost jet component measured at 8.4/8.7~GHz and 5~GHz.}
  \label{x-modelfit}
\begin{tabular}{lcc}        
\hline                 
Date        & Frequency  &  Separation \\
            & (GHz)      &  (mas)  \\
\hline                 
2005 Jul 09 & 8.6        & $ 1.06\pm 0.03$ \\
2008 Nov 13 & 5.0        & $ 1.35\pm 0.02$ \\
2012 Feb 20 & 8.4        & $ 1.48\pm 0.04$ \\
2012 Mar 09 & 8.4        & $ 1.14\pm 0.14$ \\
2014 Aug 06 & 8.7        & $ 2.26\pm 0.03$ \\
2017 Jul 09 & 8.7        & $ 2.42\pm 0.04$ \\
2018 Jul 01 & 8.7        & $ 2.35\pm 0.02$ \\
\hline
\end{tabular}
\end{table}

\subsection{Total flux density monitoring}
\label{ovro}

\begin{figure*}
\begin{center}
\includegraphics[scale=0.8]{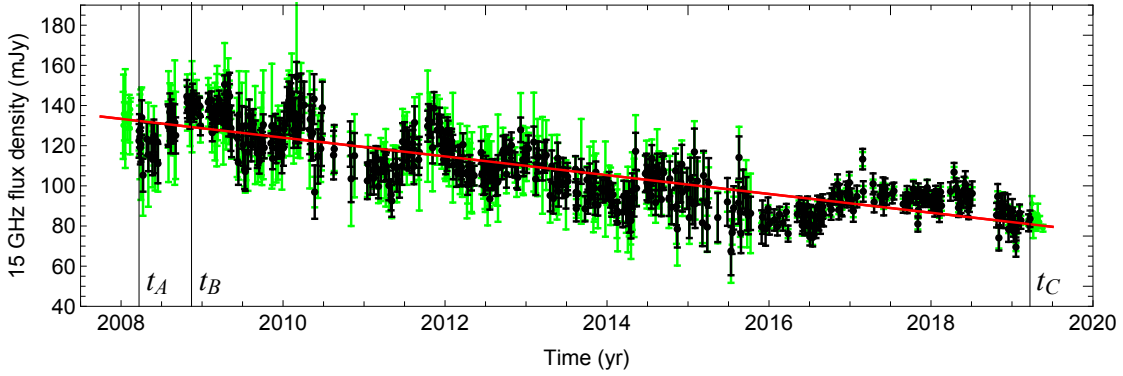}
\end{center}
\caption{The OVRO single-dish flux density curve of Spikey at 15~GHz. The original measurement points with error bars are shown in green, overlaid by the smoothed data (i.e. the data points with re-scaled error bars) in black. The best-fit linearly decreasing trend is indicated by the red line (see Sect.~\ref{ovrofluxden}). The labels $t_\mathrm{A}$ (at 2008.222) and $t_\mathrm{C}$ (at 2019.222) mark the first and last epochs of the smoothed flux density curve, while $t_\mathrm{B}$ marks the epoch of the 1.7- and 5-GHz VLBI observations (2008.870).}
\label{fig:ovrofluxden}
\end{figure*}

The quasar J1918+4937 is included in the sample of extragalactic sources regularly monitored with the 40-m OVRO radio telescope at 15~GHz frequency \citep{richards11}. The total flux density variability of J1918+4937 from early 2008 to date can be seen in Fig.~\ref{fig:ovrofluxden} which we constructed from the monitoring data available at the OVRO website\footnote{\url{http://www.astro.caltech.edu/ovroblazars/}}. A few flux density data points ($\sim9$\% of the total number) with excessively large error bars were discarded as unreliable. 

The data after 2015 November 28 (2015.9) have error bars typically a factor of $\sim 2$ smaller than before. This is likely due to the new receiver installed at OVRO in 2014 June, and a new data processing pipeline used. Keeping in mind that the time sampling of the flux density curve is more or less uniform, we re-scaled the error bars, in order to associate comparable weights to the older and the more recent data for a subsequent model fitting. The procedure was as follows. For the $n$-th data point, we calculated the standard deviation of the flux densities in the range $[(n-k)\ldots (n+k)]$ with $k=10$, and assigned it to the $n$-th data point as its new error bar. This way, the first and last $k$ data points had to be dropped from the light curve. The 15-GHz flux densities with the smoothed error bars are shown in Fig.~\ref{fig:ovrofluxden}, overlaid on the original light curve.

\section{Results}
\label{jetmodel}

\subsection{Jet parameters}
\label{jetparam}

Figure~\ref{images} shows an asymmetric radio structure with a compact core and a one-sided extension. It is typical for bright radio-loud quasars where the emission from one of the intrinsically symmetric jets that is pointing close to the observer's line of sight is enhanced by relativistic beaming \citep[for a recent review, see][]{blandford19}. In the case of J1918+4937, the approaching jet is pointing towards the Northwest as projected on the sky. The radio emission can be traced out to about 80~mas at the lower observing frequency, 1.7~GHz, then it becomes diffuse and resolved out on the long interferometer baselines. This angular extent corresponds to a projected linear size of 630~pc.

The bright VLBI core at the southeastern end of the nearly straight structure (Fig.~\ref{images}) is in fact the base of jet where it becomes optically thick at the given observing frequency. The fitted Gaussian model parameters of the core (Table~\ref{modelfit}) can be used to calculate the apparent brightness temperature, 
\begin{equation}
T_{\rm b} = 1.22\times 10^{12}\frac{F}{\theta ^{2}\nu ^{2}}(1+z_\mathrm{sp}) \,\, {\rm K}, 
\label{equ1}
\end{equation}
where $F$ is the  flux density measured in Jy, $\theta$ the diameter of the circular Gaussian component in mas (full width at half-maximum, FWHM), and $\nu$ the observing frequency in GHz. Taking into account the redshift of J1918+4937, $z_\mathrm{sp}=0.926$, the core brightness temperatures are $(2.6 \pm 0.4) \times 10^{11}$~K and $(2.9 \pm 0.2) \times 10^{11}$~K at 1.7 and 5~GHz, respectively. These values agree within their uncertainties, so we adopt $T_{\rm b} \approx 2.7 \times 10^{11}$~K for the further calculations.

The ratio between the apparent and the intrinsic brightness temperatures gives the Doppler-boosting factor, $\delta = T_{\rm b} / T_{\rm b, int}$. If we follow the usual practice and assume the equipartition brightness temperature \citep{readhead94} as $T_{\rm b,int} \approx 5 \times 10^{10}$~K, then the Doppler factor is $\delta \approx 5$. On the other hand, based on measurements of a sample of pc-scale jets, \citet{homan06} arrived at a somewhat lower typical intrinsic brightness temperature value, $T_{\rm b,int} \approx 3 \times 10^{10}$~K. Considering this, the Doppler factor of the jet in J1918+4937 would become $\delta \approx 9$.   

The amount of Doppler boosting depends on two fundamental jet parameters, the bulk Lorentz factor ($\Gamma$) of the plasma flow (i.e. the intrinsic jet speed) and the jet inclination with respect to the line of sight ($\iota_0$). If the apparent proper motion of the jet components can be measured based on VLBI imaging observations conducted at multiple epochs, it is possible to estimate values of $\Gamma$ and $\iota_0$ as well \citep[e.g.][]{urry95}. Even though sensitive imaging data are found in the archives for J1918+4937 at a single epoch only at the above frequencies (1.7 and 5 GHz), from the available multi-epoch snapshot 8-GHz VLBI observations we were able to track the motion of one of the inner jet components. Assuming a linear outward motion (Fig.~\ref{fig:propermotion}), we estimate its apparent speed in the units of the speed of light ($c$) as $\beta_\mathrm{app} = 5.33 \pm 0.65$. If we consider $\beta_\mathrm{app}$ as a representative estimate of the apparent jet speed in J1918+4937, and take the possible values of the Doppler factors derived above, we can obtain \citep[see e.g.][]{urry95} the bulk Lorentz factor
\begin{equation}
\Gamma =  \frac{\beta_\mathrm{app}^2 + \delta^2 +1}{2 \delta}
\end{equation}
and the jet inclination angle 
\begin{equation}
\cos \iota_0 =  \frac{\Gamma - \delta^{-1}}{\sqrt{\Gamma^2 - 1}}.
\end{equation}
For $\delta = 5$, we get $\Gamma \approx 5.4$ and $\iota_0 \approx 11\fdg5$, and for $\delta = 9$, we get $\Gamma \approx 6.1$ and $\iota_0 \approx 5\fdg6$.

\subsection{Jet structural model utilizing eccentric SMBH orbit}
\label{jet-model}

While the jet shape in Fig.~\ref{images} seems remarkably straight on scales of several tens of mas, some wiggling is also apparent, especially at 5~GHz where the angular resolution is higher. Here we build up a structural (morphological) model of the jet as seen projected onto the plane of the sky, based on the fitted circular Gaussian model component positions (Table~\ref{modelfit}). We assume that these compact radio components were launched by a jetted supermassive black hole (SMBH) moving along an eccentric orbit in the binary system, and the jet launching is affected by the periodically changing orbital velocity of the jet emitter SMBH.
This idea was applied earlier in several studies \citep{Roos1993,Kun2014,Kun2015} but for circular orbits. Here we further develop the model, to allow for eccentric binary orbits with arbitrary spin angles. Note that in the jet model below, the jet components themselves move along ballistic trajectories and not along helical paths. Rather we see a helical pattern on the sky formed by the subsequently emitted components, as the angle of the jet launching changes periodically. We assume that this pattern motion preserves the jet launching angle at least up to tens of mas from the central engine. Meanwhile, the physical distances between the components are growing as the time passes.

\begin{figure}
\centering
\includegraphics[bb=260 170 900 750, scale=0.47, clip=]{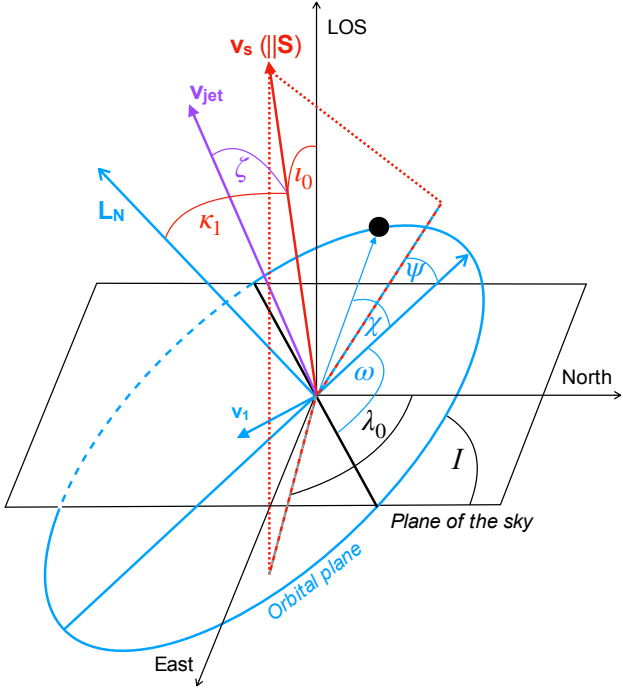}
\caption{Geometric configuration of the Spikey system centred on its barycentre. The black dot marks the position of the jet-emitting SMBH along its elliptical orbit. LOS indicates the line of sight, and $L_\mathrm{N}$ is the Newtonian orbital angular momentum. The true anomaly is $\chi$, the argument of the periapsis is $\omega$, the orbital inclination is $I$, the BH spin angle with respect to the orbital normal is $\kappa_1$, the angle between the projection of the spin onto the orbital plane and the periapsis line is $\psi$, and the inclination angle of the spin with respect to the LOS is $\iota_0$. The position angle of the spin projected onto the plane of the sky ($\lambda_0$) is measured from North through East. Furthermore, $\mathbf{v_1}$ is the orbital velocity vector of the jet-emitting SMBH at the instant of the jet component launching (if the secondary BH emits the jet, for its argument of periapsis $\omega_2=\omega_1+\pi$ holds in radians), $\mathbf{v_s}$ is the original jet velocity vector (that is parallel to the spin) and $\mathbf{v_\mathrm{jet}}$ is the vectorial sum of the above two. Finally, $\zeta$ is the instantaneous half-opening angle of the jet. For the sake of clarity, we shifted the jet velocity vector to the barycentre. In reality, the jet launches from the immediate vicinity of the emitting SMBH.}
\label{fig_bbabra}
\end{figure}

Let us assume an orthogonal coordinate system $\mathcal{K}$ in which the $z$ axis is parallel to the orbital angular momentum $\mathbf{L_\mathrm{N}}=L_\mathrm{N} \mathbf{\hat L_\mathrm{N}}$ ($z||\mathbf{\hat L_\mathrm{N}}$) (here $\mathbf{\hat L_\mathrm{N}}$ denotes the unit vector pointing to the direction of the orbital angular momentum), and the $x$ axis is directed towards the pericentre of the orbit. The orbital configuration is depicted in Fig.~\ref{fig_bbabra}. The instantaneous orbital velocity vector of the $i$-th BH in the orbital plane as a function of the eccentric anomaly $E$ is
\begin{gather}
\mathbf{v_i}(E)
 =  \begin{pmatrix}
   v_{i,x}(E) \\
   v_{i,y}(E)
   \end{pmatrix}
   =
 v_{0,i}
  \begin{pmatrix}
   -\sin \chi (E) \\
   e+\cos \chi(E)
   \end{pmatrix},
   \label{eq:v1}
\end{gather}
where $v_{0,i}$ is the circular orbital speed of the jet emitter SMBH ($i=1$ for the dominant, and $i=2$ for the secondary-mass BH),
\begin{align}
\chi (E)=2\arctan \left[\sqrt{\frac{1+e}{1-e}}\tan \frac{E}{2}\right]
\end{align}
is its true anomaly,
\begin{align}
a=\left[ \frac{G(m_1+m_2)}{4 \pi^2} T^2 \right]^{1/3}
\end{align}
is the semi-major axis of the orbit, $G$ is the gravitational constant, $T$ is the orbital period, $m=m_1+m_2$ is the total mass, and $e$ is the orbital eccentricity. If the dominant BH is the jet emitter, then its velocity should be considered in Eq.~\ref{eq:v1}, which is
\begin{align}
v_{0,1}=\frac{2\pi}{T}\frac{a}{\sqrt{1-e^2}}\frac{m_2}{m_1+m_2},
\end{align}
and if the secondary BH is the jetted one, its velocity is
\begin{align}
v_{0,2}=\frac{2\pi}{T}\frac{a}{\sqrt{(1-e^2)}}\frac{m_1}{m_1+m_2}.
\end{align}

The direction of the jetted BH spin $\mathbf{S_i}$ in $\mathcal{K}$ is the unit vector
\begin{align}
\mathbf{\hat{S}_i}=(\sin \kappa_i \cos \psi_i,\sin \kappa_i \sin \psi,\cos \kappa_i), 
\end{align}
where $\kappa_i= \arccos (\mathbf{\hat{S}_i} \cdot \mathbf{\hat{L}_N})$ is the angle between $\mathbf{S_i}$ and the orbital angular momentum $\mathbf{L_N}$, and $\psi_i$ is the angle between the projection of the spin onto the orbital $(x,y)$ plane and the periapsis line. We assume that one of the two BHs emits the jet via the Blandford--Znajek mechanism \citep{BlandfordZnajek1977}. In this case, the jet symmetry axis is directed along the BH spin $\mathbf{S_i}$, consequently the unperturbed jet velocity vector becomes $\mathbf{v_\mathrm{s}}=v_\mathrm{s} \mathbf{\hat{S}_i}$ in $\mathcal{K}$, and its components are
\begin{gather}
\mathbf{v_s}=
\begin{pmatrix}
v_{\mathrm{s},x}\\ 
v_{\mathrm{s},y}\\ 
v_{\mathrm{s},z} 
\end{pmatrix}%
=
\begin{pmatrix}
v_\mathrm{s} \sin \kappa_i \cos \psi_i\\ 
v_\mathrm{s} \sin \kappa_i \sin \psi_i \\ 
v_\mathrm{s} \cos \kappa_i%
\end{pmatrix}.
\end{gather}
 The jet velocity vector $\mathbf{v_{jet}}$ is the vectorial sum of the unperturbed jet velocity vector $\mathbf{v_{s}}$ and the orbital velocity $\mathbf{v_i}$, such that
\begin{gather}
\mathbf{v_\mathrm{jet}}=
\begin{pmatrix}
 v_{\mathrm{jet},x}\\ 
 v_{\mathrm{jet},y}\\ 
 v_{\mathrm{jet},z} 
\end{pmatrix}%
=%
\begin{pmatrix}
 v_\mathrm{s} \sin \kappa_i \cos \psi_i-v_{0,i} \sin \chi\\ 
 v_\mathrm{s} \sin \kappa_i \sin \psi_i+v_{0,i} (e+\cos  \chi)\\ 
 v_\mathrm{s} \cos \kappa_i
\end{pmatrix}.
\end{gather}

Let $\zeta$ be the angle between $\mathbf{v_{jet}}$ and $\mathbf{v_s}$, which is calculated as
\begin{align}
\sin \zeta=\frac{|\mathbf{v_{jet}}\times \mathbf{v_s}|}{|\mathbf{v_{jet}}| |\mathbf{v_s}|},
\label{eq:crosszeta}
\end{align}
where
\begin{align}
|\mathbf{v_{jet}}\times \mathbf{v_s}|=
v_{0,i} v_\mathrm{s} \cos \kappa_i \sqrt{C_1+C_2 \tan^2 \kappa_i}
\end{align}
with
\begin{align}
C_1&=1+e^2+2e\cos\chi, \nonumber\\
C_2&=(\cos(\chi-\psi_i)+e\cos \psi_i)^2, \nonumber
\end{align}
and
\begin{equation}
|\mathbf{v_{jet}}| |\mathbf{v_s}|=v_\mathrm{s} \sqrt{C_3 + C_4 + C_5}
\end{equation}
with
\begin{align}
C_3&=v_\mathrm{s}^2 \cos \kappa_i^2, \nonumber\\
C_4&=(v_{0,i} \sin \chi -  v_\mathrm{s} \cos \psi_i \sin \kappa_i)^2, \nonumber\\
C_5&=(v_{0,i} (e + \cos \chi) +   v_\mathrm{s} \sin \kappa_i \sin \psi_i)^2. \nonumber
\end{align}

For the orbital velocities in Spikey, even at this sub-pc separation, $v_{0,i} \ll v_\mathrm{s}$, and then the series expansion of their ratio (Eq. \ref{eq:crosszeta}) in leading order gives
\begin{align}
&\sin \zeta =\frac{v_{0,i} \cos \kappa_i}{v_\mathrm{s}} \times\nonumber\\
&\times \sqrt{ 1+e^2+2e \cos \chi +(\cos(\chi-\psi_i)+e\cos \psi_i)^2 \tan^2 \kappa_i} .
\end{align}

Now let us define a new orthogonal coordinate system $\mathcal{K'}$, such that its $z'$ axis is parallel to the spin of the jetted BH. In this system, the jet morphological model turns to
\begin{align}
x'(u)&=\frac{B}{2\pi}u[-\sin \chi(u-\phi)],\\
y'(u)&=\frac{B}{2\pi}u[e+\cos \chi(u-\phi)],\\
z'(u)&=\frac{A} {2\pi}u,
\end{align}
where $u$ is the polar angle \citep{Kun2014}, $\phi$ is the initial phase of $u$, and $B$ is the jet growth in mas perpendicular to its symmetry axis while $u$ changes by $2\pi$ over the time period $T_u$. This latter quantity is measured in the observer's frame as
\begin{equation}
B'=v_{0,i} \cos \kappa_i \frac{T_u}{s}(1+z_\mathrm{sp}),
\label{eq:Bgrowth}
\end{equation}
where $v_{0,i} \cos \kappa_i$ is the orbital velocity  perpendicular to $\mathbf{S}_i$, $s$ is the scale factor that relates projected linear size to the measured angular size (in pc\,mas$^{-1}$). Another parameter, $A$ is the jet growth in mas parallel to its symmetry axis while $u$ changes by $2\pi$ over the time period $T_u$. The quantity $A$ is measured in the observer's frame as
 \begin{flalign}
A'=\left( v_\mathrm{jet}+v_{0,i}\sqrt{1+e^2+2e\cos \chi(u-\phi) \cos \kappa_i} \right) \frac{T_u}{s}(1+z_\mathrm{sp}).
\label{eq:aprime}
\end{flalign}
Then we define a coordinate system $\mathcal{K''}$, such that the $x''$ and $y''$ axes point to East and North in the plane of the sky, respectively, and the $z''$ axis coincides with the direction of the line of sight (LOS), as shown in Fig.~\ref{fig_bbabra}. The inclination angle between the LOS and the spin of the jet emitter BH is $\iota_0$ (which we call spin inclination angle), and $\lambda_0$ is its position angle measured from North ($y''$ axis) through East ($x''$ axis). Employing the same rotational matrices as in \citet{Kun2014},
\begin{align}
\label{eq:jetmodelx}
x''(u)&=\left[x'(u)\cos \iota_0+ z'(u)\sin \iota_0 \right]\cos \lambda_0-y'(u)\sin \lambda_0, \\
y''(u)&=\left[x'(u)\cos \iota_0+ z'(u)\sin \iota_0\right]\sin \lambda_0+y'(u)\cos \lambda_0.
\label{eq:jetmodely}
\end{align}
 In this model, the helical jet shape is in fact the pattern drawn by the perturbed jet components ejected at different epochs from the central engine. In other words, the individual components do not move along a helix, rather the pattern they collectively form grows both in the direction of the spin and perpendicular to it, as described by the parameters $A$ and $B$, respectively.

\subsection{Application of the jet model to the VLBI data}
\label{jet-model-appl}

\begin{figure*}
\begin{center}
\includegraphics[scale=0.47]{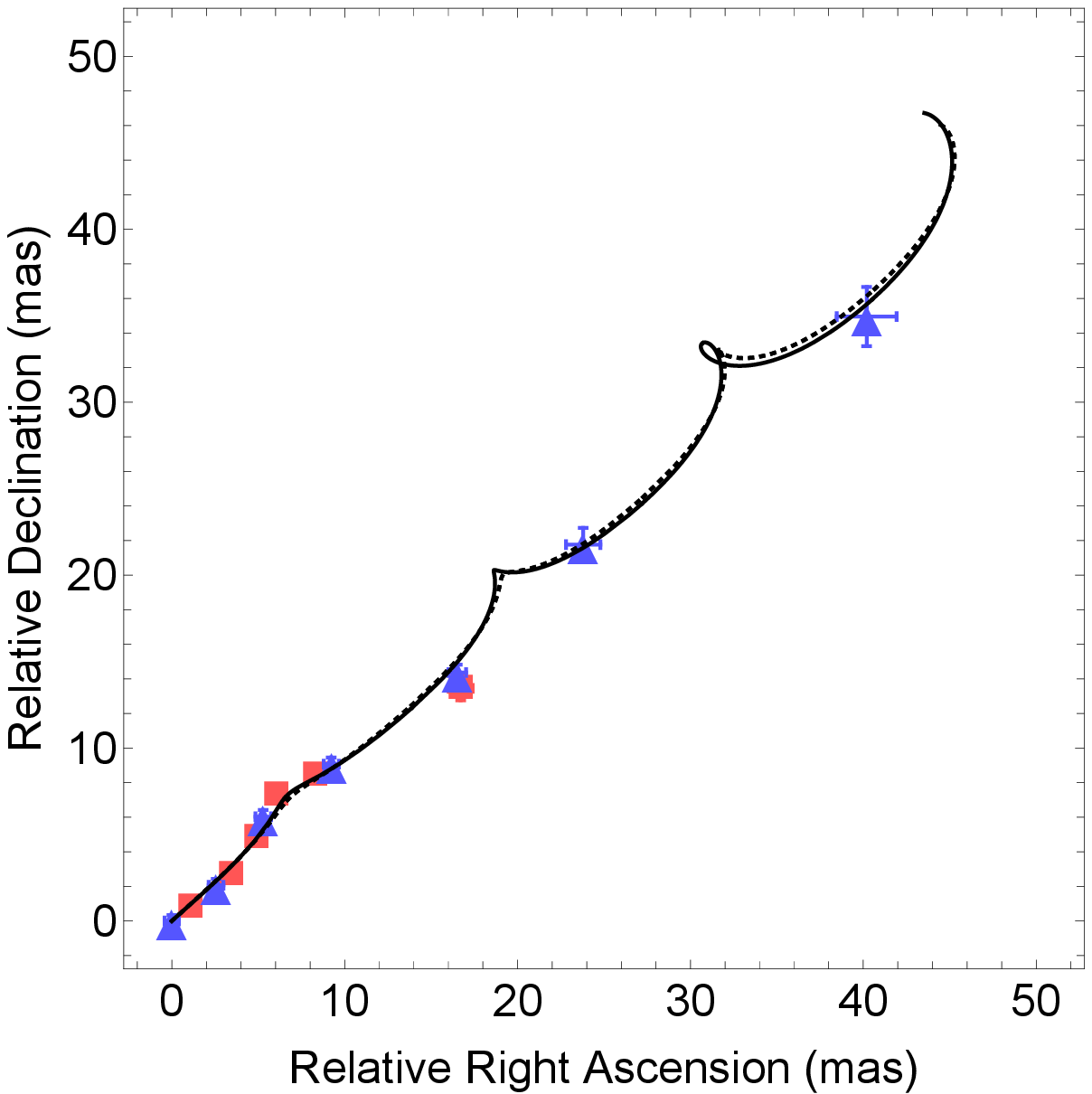}
\includegraphics[scale=0.48]{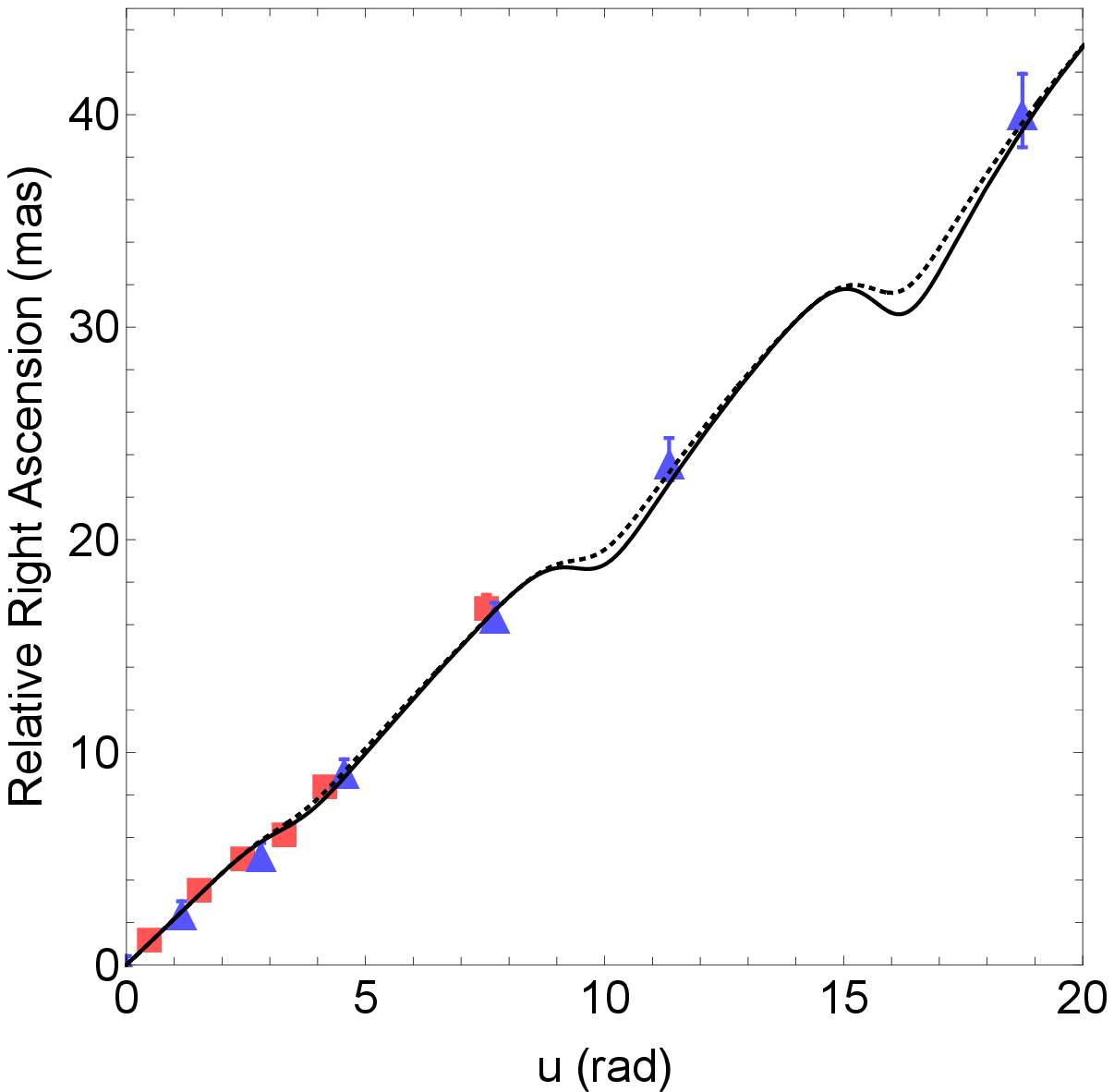}
\includegraphics[scale=0.48]{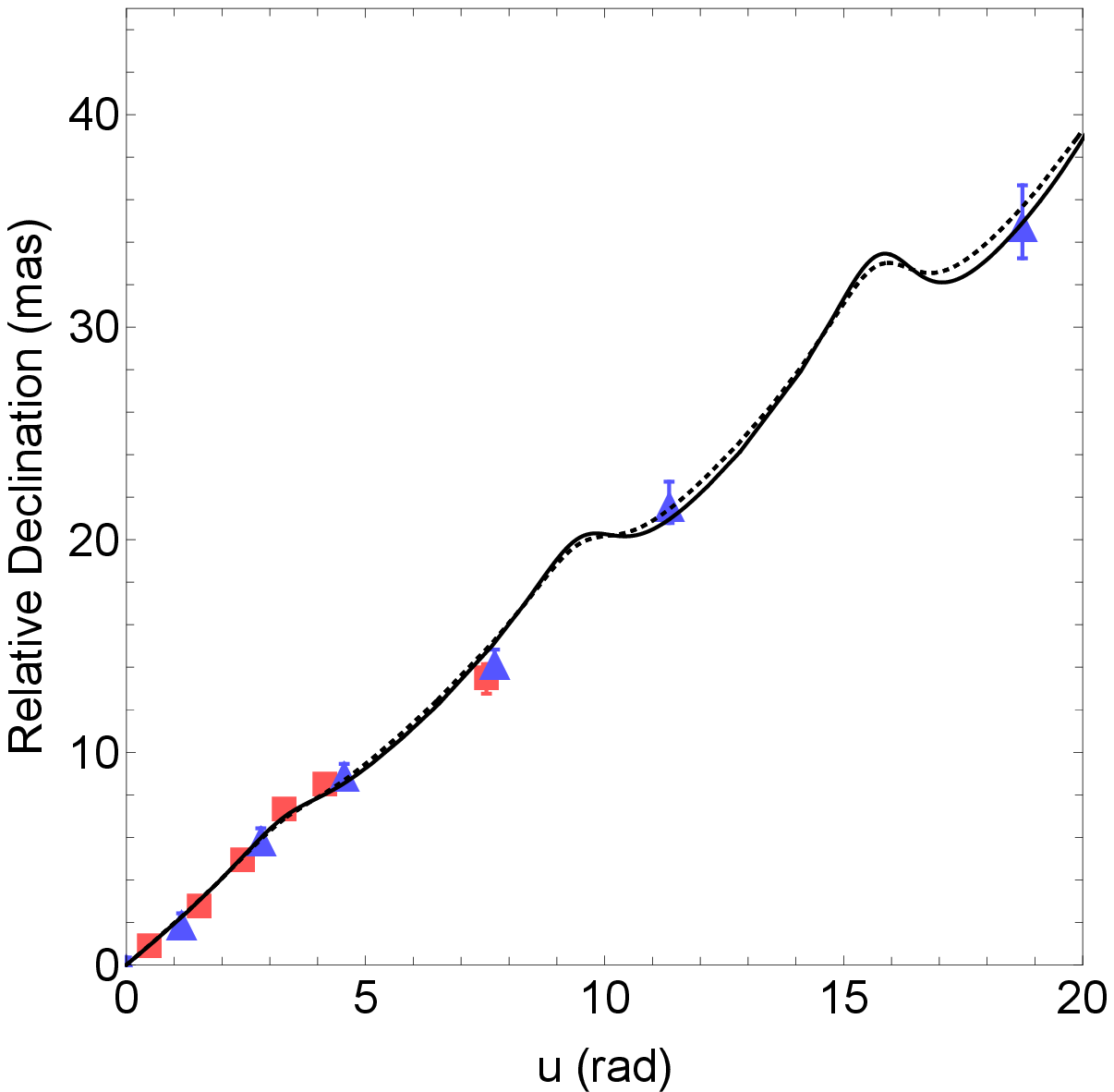}
\end{center}
\caption{The modeled jet shape fitted to the 1.7- and 5-GHz VLBI component positions, marked by blue triangles and red squares, respectively. \textit{Left:} the right ascension and declination of the components relative to the VLBI core and the best-fit jet shape assuming Doppler factor $\delta=5$, once with the dominant-mass ($m_1$) SMBH being the jet emitter (dotted black curve) and then with the secondary-mass ($m_2$) SMBH (continuous black curve). \textit{Middle:} the right ascension of the components relative to the VLBI core as a function of $u$. \textit{Right:} the declination of the components relative to the VLBI core as a function of $u$. The jet structure is rotated by $90\degr$ towards East.}
\label{fig:bestfitd5}
\end{figure*}
\begin{table*}
  \caption[]{Grid parameters of the best-fit models, such as projected pitch along the spin ($A''$), Lorentz factor ($\Gamma$), spin angle ($\kappa$), and those derived from them, the spin inclination angle ($\iota_0$) and the jet speed $\beta$, assuming that either the dominant-mass SMBH \textit{(top)} or the secondary SMBH \textit{(bottom)} launches the jet. We show the lowest $\chi$-square values ($\chi^2_\mathrm{min}$). We also list the parameter ranges in which the models are indistinguishable from each other (i.e. $\Delta \mathrm{AIC}\leq2$) and give the averages and standard deviations of the parameters in those ranges.}
  \label{modelparsm12}
\begin{tabular}{c||cc|cc}
\hline 
\multicolumn{5}{c}{The jet is emitted by the larger mass SMBH ($m_1$)} \\ 
\hline 
 & \multicolumn{2}{c}{$\delta=5$ ($\chi^2_\mathrm{min}=34.54$)} & \multicolumn{2}{|c}{$\delta=9$ ($\chi^2_\mathrm{min}=33.84$)} \\
\hline 
 &  \multicolumn{2}{c}{$\Delta \mathrm{AIC}\leq2$} &   \multicolumn{2}{c}{$\Delta \mathrm{AIC}\leq2$}\\ 
\hline 
$A''$(mas)&  [16.6:19.3] & $ 17.9 \pm 0.6 $ &[16.7:19.1] & $ 17.9 \pm 0.6 $\\
$\Gamma$&  [3.0:20.0] & $ 16.2 \pm 4.9 $  & [5.0:20.0] & $ 12.7 \pm 4.2 $\\
$\kappa$($\degr$)& [0:30] & $ 11.0 \pm 7.5 $& [0:64] & $ 23.5 \pm 13.2 $\\
$\iota_0$($\degr$)& [7.6:8.8] & $ 8.0 \pm 0.3 $ & [3.7:6.4] & $ 5.9 \pm 0.4 $\\
$\beta$($c$)&  [0.943:0.999] & $0.992\pm 0.017$  & [0.980:0.999] & $ 0.992 \pm 0.008 $\\
\hline 
\multicolumn{5}{c}{The jet is emitted the smaller mass SMBH ($m_2$)} \\ 
\hline 
 & \multicolumn{2}{c}{$\delta=5$ ($\chi^2_\mathrm{min}=31.42$)} & \multicolumn{2}{|c}{$\delta=9$ ($\chi^2_\mathrm{min}=31.85$)} \\
\hline 
 &  \multicolumn{2}{c}{$\Delta \mathrm{AIC}\leq2$} &  \multicolumn{2}{c}{$\Delta \mathrm{AIC}\leq2$}\\ 
 \hline 
$A''$(mas)& [16.7:19.1] & $ 17.9 \pm 0.6 $ &  [16.7:19.1] & $ 17.9 \pm 0.6 $\\
$\Gamma$& [3.0:20.0] & $ 10.6 \pm 4.9 $ &  [5.0:20.0] & $ 12.4 \pm 4.3 $\\
$\kappa$($\degr$)& [65:79] & $ 73.0 \pm 3.0 $ & [77:85] & $ 79.8 \pm 1.4 $\\
$\iota_0$($\degr$)& [7.6:11.5] & $ 9.7 \pm 1.3 $ &  [3.7:6.4] & $ 5.9 \pm 0.5 $\\
$\beta$($c$)& [0.943:0.999] & $ 0.990 \pm 0.012 $ & [0.980:0.999] & $ 0.995 \pm 0.004 $\\
\hline
\end{tabular} 
\end{table*}
After setting up the model to describe the jet structure, we now take into account the measured VLBI component positions at 1.7 and 5~GHz (Table~\ref{modelfit}) and derive the model parameters. These observations were made at the same time, but at different frequencies, and the position of the optically thick core components (i.e., the base of the jet used as a reference for the relative position of other components further along the jet) is known to depend on the observing frequency, an effect called core shift \citep{Blandford1979,lobanov98}.
\citet{Sokolovsky2011} conducted a dedicated survey with the VLBA at nine frequencies in the $1.4-15.4$~GHz range to quantify the core-shift effect in $20$ AGN jets. The average (and median) core shift between $1.7$ and $5$~GHz was found to be approximately 0.9~mas. This is comparable to the uncertainties of our component positions (Table~\ref{modelfit}). Therefore we used model components fitted at both 1.7 and 5~GHz together in the further analysis. Note that the angular resolution of the interferometer is about 3 times better at the higher frequency. Thus the inner section of the jet is characterised by more components at 5~GHz, while the outer section is only seen at 1.7~GHz, where the array is more sensitive to the weaker, extended, steep-spectrum features.

We are in a unique situation because some of the key parameters of the Spikey SMBHB system are accurately known from \citet{hu20}. Therefore we adopt the (rest-frame) orbital period $T=1.14$~yr, the orbital inclination $I=1.43$~rad, the mass of the primary SMBH $m_1=2.5\times10^7 \mathrm{M}_\odot$, the mass of the secondary SMBH $m_2=5.0\times10^6 \mathrm{M}_\odot$, and the orbital eccentricity $e=0.52$. These numbers imply $a=5.1\times10^{13} \mathrm{m} = 0.002$~pc and the circular velocities $v_{0,1} \approx0.006 c$ and $v_{0,2}\approx 0.03 c$. The bulk jet speed (expressed in the units of $c$) and the spin inclination angle with respect to the LOS are
\begin{align}
\beta_\mathrm{s}=\sqrt{1-\frac{1}{\Gamma^2}}
\end{align}
and
\begin{align}
\iota_0=\arccos \left\{ \frac{1}{\beta_\mathrm{s}} \left(1-\frac{1}{\Gamma \delta}\right) \right\},
\end{align}
respectively \citep[e.g.][]{urry95}, where $\beta_\mathrm{s}=v_\mathrm{s}/c$.

We apply non-linear least squares curve (parametric) fitting with $\sigma^{-2}$ weights by employing the Levenberg--Marquardt algorithm to get the best-fit jet model, such that the $\chi^2$ was minimized during the process. As a next step, we characterise the reliability of our best-fit model and investigate whether there are other solutions that cannot be discriminated from the above one, solely based on their $\chi^2$ value. The Akaike information criterion \citep[AIC, ][]{Akaike1974} estimates the quality of each model relative to each of the others, i.e. it is a tool for model selection, either for nested or not nested models. The lower the AIC, the better the performance of the given model. Models in which the difference in AIC relative to $\mathrm{AIC}_\mathrm{min}$ is $\leq2$ perform approximately equally \citep{Burnham2002}, therefore the selection of any of them might lead to inconclusive statements. Here, as the number of parameters is the same, we select the models of approximately equal quality solely based on their $\chi^2$ values. 

If we apply the Doppler factor $\delta=5$ (see Sect.~\ref{jetparam}), then a lower limit for the Lorentz factor is $\Gamma_\mathrm{min}=2.6$. This corresponds to the case when the jet is seen exactly pole-on (i.e. $\iota_0=0$). For a numerical parameter estimation, we set up a grid where the projected jet growth along the spin direction, $A''=A' \sin \iota_0$ changes from $10$ to $30$~mas (in steps of $\Delta A''=0.1$~mas), $\Gamma$ changes from $3$ to $20$ (in steps of $\Delta \Gamma=0.5$), and $\kappa_i$ changes from $1\degr$ to $90\degr$ (in steps of $\Delta \kappa_i=1\degr$). The bulk jet speed varies from $0.9428\,c$ to $0.9987\,c$ on the grid as $\Gamma$ changes between $3$ and $20$. The only parameter we have to solve for is $\phi_i$, while $A''$, $\Gamma$, and $\kappa_i$ are changing along the grid as described above. Since $v_{0,i}\ll v_\mathrm{jet}$, we neglect the term corresponding to $v_{0,i}$ in $A'$ (Eq.~\ref{eq:aprime}), and the jet grows along the spin direction solely as a result of the non-zero jet velocity $v_\mathrm{jet}$.

By fitting the jet model described by Eqs. \ref{eq:jetmodelx}-\ref{eq:jetmodely}, the following best-fit parameters emerged if we assume the dominant-mass BH as the jetted one: $A''=17.9$~mas, $\Gamma=20.0$, $\kappa_1=0\degr$ (with the lowest $\chi^2=34.54$, reduced $\chi_\mathrm{R}^2=1.38$). The modeled jet shape corresponding to these values and the measured VLBI component positions are plotted in Fig.~\ref{fig:bestfitd5}.
After considering the best-quality models leading to $\mathrm{AIC}$ difference from $\mathrm{AIC}_\mathrm{min}$ as $\Delta\mathrm{AIC} \leq 2$, we calculate the average value and the standard deviation of the grid parameters. We repeated the process with Doppler factor $\delta=9$ (corresponding to $T_\mathrm{b,int}=3\times10^{10}$ K; see Sect.~\ref{jetparam}). Selecting the best-quality models, we get again the average value and the standard deviation of their grid parameters. The best-fit grid parameters, as well as parameters of models giving the same performance are summarized in Table \ref{modelparsm12} for $\delta=5$ and $\delta=9$. We also show here the spin inclination angles ($\iota_0$) and jet speeds ($\beta$) derived from the corresponding grid parameters. It seems that the model is not very sensitive to the Lorentz factor, which is not surprising because the same projected jet opening angle can be generated with a variety of parameter pairs if we allow to simultaneously change the jet growth in the direction to the spin and perpendicular to it. Note that the best-fit jet structure model ($\chi^2=34.54$) is achieved with $\kappa_1=0\degr$, and the parameter range of $\kappa_1$ giving models with comparable quality emerged as [$0\degr$:$30\degr$]. The orbital velocity of the more massive SMBH is relatively small compared to the jet velocity along the spin because of the small BH mass ratio in Spikey. The fitting process tries to balance it with increasing the $\cos \kappa_i$ term in Eq.~\ref{eq:Bgrowth} in order to model the observed jet growth perpendicular to the spin as closely as possible.

We repeated the jet-shape-fitting process, now assuming the $m_2$ mass SMBH as the jetted one. The grid parameters of the best-fit jet model, the parameter ranges in which the models lead to $\Delta \mathrm{AIC}\leq2$, as well as the averages and standard deviations of the parameters in these ranges are summarized in Table \ref{modelparsm12}. The modeled jet shape corresponding to these values is plotted in Fig.~\ref{fig:bestfitd5}. If the secondary-mass SMBH is assumed as the jet emitter, then the best-fit model gives $\kappa_2=73\degr$ if $\delta=5$, with a slightly lower $\chi^2=31.42$ compared to the value found for the primary SMBH. The case is similar for $\delta=9$, and the corresponding parameter ranges leading to $\Delta \mathrm{AIC}\leq2$ are much more tightly constrained, without containing the limiting $\kappa_i=0$. This is because the velocity of the secondary SMBH is much larger compared to the more massive one, and the observed jet growth perpendicular to the spin can be modeled without maximizing the $v_{0,i} \cos \kappa_i$ term in Eq.~\ref{eq:Bgrowth}.

\subsection{Total flux density variations}
\label{ovrofluxden}

The observed period in the optical light curve of Spikey is $T_\mathrm{obs}=(1+z_\mathrm{sp}) T=805$~d, which was recognised as the observed orbital period of the SMBH binary \citep{hu20}. The 15-GHz radio flux density curve (Fig.~\ref{fig:ovrofluxden}, Sect.~\ref{ovro}) measured at OVRO \citep{richards11} indicates a decreasing trend on a long term, together with some flaring activity, and possibly a longer flare started at around 2015 November. If we interpret the flux density changes as quasi-periodic, a signal with $500-600$-d period might seem superimposed on the linear trend. This period is $\sim 200-300$~d shorter than the one in the optical light curve and therefore we could not reliably fit a periodic component by employing the proposed binary parameters of Spikey \citep{hu20}, where the periodically strengthened Doppler boosting would readily explain the radio flux density variations. The expected periodic effect is most likely masked by the episodic activity of the jetted AGN in the system. 
Instead we fitted a simple linear function to the smoothed data (see Sect.~\ref{ovro}), resulting in a slope of $(-4.67\pm0.13)$~mJy\,yr$^{-1}$. This trend is also shown in Fig.~ \ref{fig:ovrofluxden}. Also, we cannot exclude the possibility that some level of radio emission is associated with the second SMBH component.

The decreasing trend in the flux density curve might indicate that the average inclination angle of the jet becomes larger with time. In the framework of the SMBHB model, this can be interpreted as the jet direction gradually moving away from the line of sight, therefore decreasing the Doppler boosting effect on the observed radio emission. Below we investigate whether this scenario is consistent with the known binary parameters \citep{hu20}. 

The orbital period in the order of years and the sub-pc separation in Spikey indicate that the binary has already progressed into the inspiral evolutionary phase of the merger, i.e., the third and final stage where the gravitational radiation becomes the dominant dissipative effect over dynamical friction and gravitational slingshot interactions \citep[e.g.][]{merritt05}. In the inspiral phase, the dynamical evolution of the binary can be treated analytically by expanding the equation of motion in terms of the so-called post-Newtonian (PN) parameter as follows \citep{Kidder1995}: 
\begin{flalign}
\frac{d^2\mathbf{r}}{dt^2}=-\frac{m\mathbf{r}}{r^3} \left(1+\mathcal{O}(\varepsilon)+\mathcal{O}(\varepsilon^{1.5})+\mathcal{O}(\varepsilon ^{2})+\mathcal{O}(\varepsilon ^{2.5})+...\right),
\label{eqmotion}
\end{flalign}
where $\mathbf{r}$ is the binary separation being
\begin{gather}
\mathbf{r}
 =  \begin{pmatrix}
   a \cos E-e \\
   a \sqrt{(1-e^2)}\sin E\\
   0
   \end{pmatrix}
   \label{eq:instr}
\end{gather}
in the coordinate system $\mathcal{K}$. Here $\varepsilon=Gmc^{-2}r^{-1}$ is the PN parameter with $r=a(1-e^2)(1+e\cos \chi)^{-1}$, and $\mathcal{O}%
(\varepsilon ^{n})$ represents the $n$-th PN order. For the eccentric orbit in Spikey, we average the PN parameter for one orbit:
\begin{align}
\langle \varepsilon(t) \rangle = \frac{1}{T} \int_{0}^{T} \frac{Gm}{c^2}\frac{1+e \cos \chi(t')}{a(1-e^2)} dt'\approx0.001,
\end{align}
which value suggests Spikey recently entered into its inspiral phase, where $0.001\lesssim \varepsilon \lesssim 0.1$ \citep{Gergely2009, Levin2011}.

Up to 2PN orders, the merger dynamics is conservative, the constants of motion being the total energy and the total angular momentum vector $\mathbf{J}=\mathbf{S_1}+\mathbf{S_2}+\mathbf{L}$, where $\mathbf{L}$ is the total orbital angular momentum. The SMBH spins obey precessional motion \citep{Barker1975,Barker1979}:
\begin{flalign}
\dot{\mathbf{S}}_\mathbf{i}=\mathbf{\Omega_i} \times \mathbf{S_i},
\end{flalign}
where the $i$ index refers to the first or second component of the binary. The angular velocity $\mathbf{\Omega_i}$ of the $i$-th spin $\mathbf{S_i}$ contains up to 2PN order spin--orbit (1.5PN), spin--spin (2PN), and quadrupole momentum contributions (2PN). For the typical mass ratios $\nu \in [1/30\ldots 1/3]$, only the dominant spin counts \citep{Gergely2009}. The mass ratio in Spikey is $\mathbf{\nu \approx 1/5}$, so it falls into the above range implying the second spin might be neglected in the binary dynamics. In 1.5PN, the spin--orbit precession of the spins $\mathbf{S_1}$ and $\mathbf{S_2}$ occurs with angular velocities
\begin{align}
\mathbf{\Omega_1}&=\frac{G(4+3\nu)}{2c^3 r^3} \mathbf{L_\mathrm{N}} \label{eq:omega1} \,\, \mathrm{and}\\
\mathbf{\Omega_2}&=\frac{G(4+3\nu^{-1})}{2c^3 r^3} \mathbf{L_\mathrm{N}},
\label{eq:omega2}
\end{align}
respectively, where  $\mathbf{L_N}=\mu \mathbf{r} \times \mathbf{v}$ is the Newtonian orbital angular momentum, $\mu=m_1 m_2/m$ is the reduced mass which moves with velocity $\mathbf{v}$. Employing the formulae of the instantaneous separation given in Eq.~\ref{eq:instr} and the orbital velocity vector given in Eq.~\ref{eq:v1} (both expressed in $\mathcal{K}$)
\begin{align}
&\mathbf{L_N}=-a \mu \sqrt{\frac{Gm(2 a-r)}{a r}}\times \nonumber\\
&\ \times \left( e^2+e \cos \chi-\cos E(e+\chi)-\sqrt{1-e^2}\sin E \sin \chi \right)  \mathbf{\hat{L}_N}.
\label{eq:newtorbangmom}
\end{align}

The time dependence of $r$, $\chi$, and $E$ can be given by solving the Kepler equation $E(t)-e \sin E(t)=2\pi /T (t-\tau)$, where $\tau$ is the time of pericentre passage. Substituting Eq. \ref{eq:newtorbangmom} into Eqs. \ref{eq:omega1}-\ref{eq:omega2}, and averaging the spin--orbit precession period $T_\mathrm{SO}=2 \pi {\Omega}$ over one orbit, we get a value for the dominant-mass SMBH as $\langle T_\mathrm{SO,1}(t) \rangle (1+z_\mathrm{sp})\approx 15,700$~yr and for the secondary SMBH as $\langle T_\mathrm{SO,2}(t) \rangle (1+z_\mathrm{sp})\approx 3,800$~yr in the observer's frame. 

Assuming that the bulk Lorentz factor in the jet remains constant with time, and the long-term decreasing trend in the OVRO flux density curve (Fig.~\ref{fig:ovrofluxden}) is solely due to the secular change of the spin inclination angle, we calculate the possible jet inclination angles at two epochs of the OVRO flux density monitoring period by employing the flux density ratio below:
\begin{align}
\left(\frac{F_1}{F_2}\right)^{1/3}=\frac{1-\beta_s \cos \iota_2}{1-\beta_s \cos \iota_1},
\label{eq:fluxratio}
\end{align}
where the indices $1,2$ mark the flux density and spin inclination angle at two arbitrary epochs. We assumed a flat radio spectrum.
In Fig.~\ref{fig:ovrofluxden}, we marked three different epochs, $t_\mathrm{A}$, $t_\mathrm{B}$, and $t_\mathrm{C}$, which are the starting epoch of the smoothed OVRO flux density curve, the epoch of the $1.7$- and $5$-GHz VLBA observations, and the last epoch of the smoothed OVRO flux density curve, respectively. The mean 15-GHz flux densities at these three epochs are $F(t_\mathrm{A})=132$~mJy, $F(t_\mathrm{B})=130$ ~mJy, and $F(t_\mathrm{C})=81$~mJy, respectively, based on the $(-4.67\pm0.13)$~mJy\,yr$^{-1}$ slope of the linear function fitted to the flux density data. 
Employing the minimum and maximum spin inclination angles allowed by the VLBI measurements at epoch $t_\mathrm{B}$ in the framework of the present binary model, $\iota_\mathrm{0,min}=3\fdg7$ (with $\Gamma=5$, $\delta=9$) and $\iota_\mathrm{0,max}=11\fdg5$ (with $\Gamma=5$, $\delta=5$), and the flux density ratio in Eq.~\ref{eq:fluxratio}, we calculate minimum and maximum spin inclination angles at the starting and finishing OVRO epochs. The resulting possible spin inclination angles are summarized in Table~\ref{table:iotaangles}. According to our results, the spin inclination angle could have changed by $2\fdg5$--$2\fdg6$ over $11$~yr in the framework of the present model.

By expanding the equation of motion in terms of the PN parameter, as we have seen, the dynamical evolution of the binary can be treated analytically while it progresses through the inspiral phase where $0.001\lesssim \varepsilon \lesssim 0.1$ \citep{Gergely2009, Levin2011}. The time scale of the spin-flip is proportional to $\varepsilon^{-9/2}$, while the time scale of the spin--orbit precession is proportional to $\varepsilon^{-5/2}$, when the spin is comparable to the orbital angular momentum ($S_1\approx L$). For Spikey, $\langle \varepsilon(t) \rangle \approx 0.001$ means that if the flip occurs, it happens on a time scale more than $10^6$ times longer than the time scale of the precession. We can safely state that if the slow decrease in the total flux density of Spikey is indeed due to the increase of the spin inclination angle, then the underlying mechanism should be the spin--orbit precession, not the spin-flip.

\begin{table}
\caption[]{Possible jet inclination angles at the beginning (2008.22) and at the end (2019.22) of the OVRO flux density curve, based on the estimated minimum and maximum jet inclination values at the epoch of VLBI observations (2008.87). If we assume the minimum (maximum) inclination angle at 2008.87, the inclination angle changes from $3\fdg6$ ($10\fdg9$) to $6\fdg2$ ($13\fdg4$) in 11 years of the OVRO observations. These angles are marked by boldface (italic) in the table, respectively.}
\label{table:iotaangles}
\centering
\begin{tabular}{p{0.8cm}p{0.8cm}p{0.4cm}p{0.4cm}p{0.6cm}p{0.6cm}p{0.6cm}p{0.6cm}}
\hline 
$t_1$ & $t_2$ & $F_1$ & $F_2$ & $\iota_\mathrm{min}(t_1)$ & $\iota_\mathrm{min}(t_2)$  &$\iota_\mathrm{max}(t_1)$ &  $\iota_\mathrm{max}(t_2)$  \\ 
(yr) & (yr) & (mJy)&(mJy)& ($\degr$)& ($\degr$)& ($\degr$)& ($\degr$)\\
\hline 
$2008.22$  & $2008.87$ & $132$ & $130$ & $\mathbf{3.6}$   &$3.7$&$\mathit{10.9}$  &  $11.5$  \\ 
$2008.87$  & $2019.22$ & $130$ & $81$ & $3.7$ &$\mathbf{6.2}$  &  $11.5$ &  $\mathit{13.4}$ \\ 
\hline 
\end{tabular}
\end{table}

\section{Discussion}
\label{discussion}

\subsection{No spike in the radio light curve}

The long-term 15-GHz OVRO monitoring \citep{richards11} covers the time of the {\em Kepler} spike \citep{smith18} occured in 2011 June. Since the optical flare lasted only for about 15 days, it was poorly sampled in the radio. However, there are 3 measurement points available in the OVRO data set for J1918+4937 in this time range, roughly at the beginning, middle, and end of the optical spike. From these data, there is no evidence for any radio brightening around Julian Date 2455724. On the contrary, the 15-GHz flux density stays constant within the measurement errors.

Why is the radio emission unaffected in the SMBHB self-lensing scenario that \citet{hu20} proposed for the optical spike? There are two possible reasons. First of all, if only one of the BHs powers a radio jet, and this one is the lensing object in the foreground, then a radio magnification is obviously not expected. But even if the lensed object in the background is a radio-loud AGN, an optical spike is not necessarily expected to be coupled with a radio brightening. The optical emission of AGNs is known to originate mainly from the accretion disk on the scale of $\sim 10^{-5}$~pc \citep[e.g.][]{koratkar99}. On the other hand, most of the 15-GHz radio emission comes from an ultracompact region downstream the jet, on $\sim0.1-1$~pc projected scale \citep[e.g.][]{lobanov98}. However, according to the model of \citet{hu20}, the SMBHB separation in the Spikey system is at least two orders of magnitude smaller. The entire binary system is therefore located well inside the region where the 15-GHz radio emission originates from. There is nothing to be gravitationally lensed in the Spikey system in radio, and even the superior angular resolution of VLBI is insufficient to directly resolve the companions.  

\subsection{Jet modeling with accurate binary parameters}

 Modeling the observed high-resolution structure and kinematics of VLBI jets in quasars is usually applied to infer parameters of suspected SMBH binaries \citep[e.g.][]{Britzen2001,Lobanov2005,Britzen2012,Valtonen2012,Caproni2013,Kun2014,Kun2015,Kun2018}.  In some of these cases, there is independent indication for the existence of the binary, e.g. from periodic optical variability. However, in the case of J1918+4937 (Spikey), the analysis of the {\em Kepler} light curve by \citet{hu20} offers more than simply an indication. The measured optical spike is a unique phenomenon requiring special circumstances, therefore its successful modeling with gravitational self-lensing and orbital Doppler boosting provides accurately determined BH masses, orbital parameters and geometric constraints for the system \citep{hu20}. Unlike the usual practice, these parameters could therefore be fed directly into the VLBI jet model presented here (Sect.~\ref{jetmodel}). It was necessary to refine this model to allow for highly eccentric binary orbits. In all earlier modeling, circular orbits were assumed for simplicity, as no reliable information about the binary orbital parameters were available. 
 
We used VLBI imaging data taken at 1.7 and 5~GHz frequencies for Spikey, and also investigated the long-term OVRO flux density monitoring measurements at 15 GHz in the context of the SMBHB model proposed by \citet{hu20}. The shape of the VLBI jet represented by the individual component positions is remarkably consistent with the Spikey binary parameters. The constraints we obtained on $\Gamma$ based on the single-epoch deep VLBI imaging of J1918+4937 at these two frequencies are not particularly strong (see Table \ref{modelparsm12}). Indeed, qualitatively, a jet with a given Doppler boosting factor can be produced either by relatively slowly-moving plasma blobs directed very close to the line of sight, or a fast jet with comparably larger inclination. Plausible values of $\Gamma$ and the mean jet inclination angle with respect to the LOS ($\iota_0$) could be provided only with multi-epoch VLBI jet kinematic studies \citep[e.g.][]{Lister19}.

However, utilizing also the available multi-epoch snapshot VLBI imaging observations of J1918+4937 at the 8.4/8.7-GHz frequency band, we were able to estimate the apparent speed ($\beta_\mathrm{app}$) in the jet from the measured linear proper motion of an inner jet component. The values of $\Gamma$ and $\iota_0$ derived from $\beta_\mathrm{app}$ for the two possible values of the Doppler factor ($\delta=5$ and 9) fall within the parameter ranges set by our VLBI jet stucture model based on the parameters of the orbital motion of a SMBHB along eccentric orbit \citep{hu20}. Moreover, the values estimated from jet kinematics, $\Gamma \approx 5 - 6$ and $\iota_0 \approx 6 - 12\degr$, appear more consistent with the solutions in Table~\ref{modelparsm12}, where the jet emitter is the smaller BH with mass $m_2$. 
 
As the optical emission likely arises from the gas bounded to the individual SMBHs in the binary system, the luminosity of the brighter minidisk \citep[e.g.][]{{Ryan2017}} would be Doppler boosted and this minidisk is likely the one associated with the fastest-moving secondary SMBH \citep{DOrazio2016,hu20}. The spike in the Kepler optical light curve of Spikey can be explained with the gravitational self-lensing if the larger-mass SMBH passes between the smaller-mass SMBH and the observer \citep{hu20}, magnifying the optical emission of the minidisk around the smaller SMBH. Also, \citet{hu20} successfully explained the long-term variability in the light curve of Spikey by variable Doppler boosting due to the motion of the secondary SMBH. This means that at least the smaller SMBH has an accretion disk what we see in optical. Our VLBI jet model that utilizes the binary model of \citet{hu20} is indeed more consistent with the jet parameters derived from VLBI monitoring of Spikey if we assume the secondary SMBH is the jetted one in the system. Notably, the $\chi^2$ values also indicate slightly better fits for those solutions, and if the secondary-mass SMBH is the jetted one, the parameter $\kappa_i$ is much better constrained, without reaching the limiting value $0\degr$.
 
As it is often seen in radio-loud AGNs, the OVRO monitoring light curve of Spikey (Fig.~\ref{fig:ovrofluxden}) is rather complex.  Variations with characteristic time scales of $\sim 1$~yr and shorter are superimposed on a generally decreasing trend in flux density. We attempted to relate this long-term trend seen during the entire monitoring period of more than $11$~yr to the SMBHB model in which one of the companions launches the relativistic jet and is therefore responsible for the synchrotron radio emission. Spin--orbit precession in a close binary SMBH system that is already in its inspiral phase can cause a change in the orientation of the BH spin and the jet orientation. Considering the Spikey parameters, we found that this change (about $2\fdg5$--$2\fdg6$ during $\sim 11$~yr) should have a noticeable effect during the OVRO monitoring period by driving away the jet from the line of sight and thus decreasing the Doppler boosting, effectively causing the observed gradual dimming of the radio source.
 
Based on our study, we can confidently say that the Spikey jet and the radio light curve are fully consistent with the binary SMBH model of \citet{hu20}. Both the jet shape and the long-term decreasing flux density trend can be reconciled with the proposed binary parameters and standard jet physics. However, alternative explanations cannot be excluded for the observed VLBI jet pattern, as well as for the radio light curve. Precessing jets can also be produced by tilted accretion discs around rapidly spinning BHs \citep{Liska18}, without invoking the presence of a binary companion. Any periodic or quasi-periodic effect related to the jet itself, its surrounding medium, or the jet feeding mechanism can in principle affect its observed structure. For example, plasma instabilities along the jet \citep[e.g.][]{Nakamura04} and quasi-periodic instabilities in the accretion flow \cite[e.g.][]{Honma92} can also cause wiggled jet structures. Similarly, total flux density variations can be produced by a multitude of physical effects, not only the change in the jet inclination angle. In particular, a long-term change in the bulk jet Lorentz factor could result in a similar trend seen in Fig.~\ref{fig:ovrofluxden}. The main point of why the SMBHB scenario is the most favourable one to explain the GHz VLBI jet structure of Spikey is that we already have an indication that Spikey hides a SMBHB based on the gravitational self-lensing model of \citet{hu20} and the spike seen in the optical light curve of the object.
 
\subsection{Could Spikey become a neutrino emitter AGN?}

\citet{Kun2017,Kun2019} proposed a scenario of binary SMBH evolution which naturally explains the observed high-energy (HE) neutrino emission, and leads to the emission of gravitational waves (GWs) through a sequence induced by the merger. For the typical mass ratio of merging SMBH binaries ($\nu \in [1/3 \ldots 1/30]$), $L>S_1$ is always transformed into $L<S_1$ \citep{Gergely2009}. It means the spin of the dominant BH usually flips, while spin--orbit precessing.

Three main phases of the emission of HE particles are expected in this scenario. The first one is the process of spin-flip, when the jets sweep through a large cone. The second one is after the spin-flip, when a new jet is boring into the environment, leading to the injection of more seed particles to create HE nuclei, $\gamma$-photons and neutrinos. The third one is probably in the instant of the coalescence of SMBHs, when a giant shock wave may be generated by low-frequency GWs to accelerate particles to high energies, leading to  a final burst of HE nuclei, $\gamma$-rays and neutrinos.

To speculate if Spikey could be a neutrino emitter based on the available data, it is vital to conclude which spin the jet is connected to. The jet power ($P_\mathrm{jet}$) is proportional to the mass of the central object ($m$) and the square of its dimensionless spin parameter $a^*$ \citep[e.g.][]{Narayan2012,Steiner2013}. We have seen that the optical light curve and the VLBI observations of Spikey together are slightly more consistent with the secondary SMBH being the jetted one in the system. The ratio of the spin magnitudes in Spikey is
 \begin{equation}
    \frac{S_1}{S_2}=\left(\frac{m_1}{m_2}\right)^2 \frac{a_1^*}{a_2^*}\approx 25 \frac{a_1^*}{a_2^*},
 \end{equation}
 which means the spin of the secondary SMBH might not be neglected in the binary dynamics only if its horizon rotates much faster compared to the horizon of the dominant one, i.e. if  $a_2^* \gg a_1^*$ holds for the dimensionless spin parameters. If it is the case, then the jet power would be much larger if the secondary SMBH emits the jet, because $P_\mathrm{jet}\propto (m,a^{*2})$. So the physical picture in Spikey becomes self-consistent if the horizon of the dominant-mass SMBH rotates much slower compared to the secondary SMBH. In this case, $S_2$ could be in the order of $S_1$, and eventually flip in the inspiral phase.

\section{Summary}
\label{summary}

J1918+4937 (Spikey) is so far a unique extragalactic object hosting a closely-separated ($\sim 0.002$~pc) SMBHB system where the masses of the companions, as well as the orbital and geometric parameters could be accurately determined from a narrow spike in its {\em Kepler} optical light curve, using a combined  gravitational self-lensing and orbital Doppler boosting model \citep{hu20}. At least one of the SMBH companions is a radio-loud AGN with a prominent relativistic plasma jet. Archival high-resolution radio interferometric imaging observations made with the VLBA at $1.7$ and $5$~GHz \citep{kharb10} allowed us to study its structure. We estimated the Doppler boosting caused by the small inclination angle of the jet to the line of sight. We then set up a model describing a jetted SMBH in a binary system with eccentric orbit, and investigated whether the apparently helical jet shape is consistent with the binary parameters derived for Spikey \citep{hu20}. By successfully applying our structural model to Spikey, we could derive the jet Lorentz factor and viewing angle, albeit with loose constraints. A comparison with the jet parameters inferred from multi-epoch VLBI monitoring data at 8.4/8.7~GHz, together with the somewhat better fits provided by the jet structural model suggest that the smaller-mass ($m_2$) component of the binary might be the jet-emitting BH.

We also studied the long-term single-dish 15-GHz flux density curve \citep{richards11}. While spikes similar to the optical one are not expected in the radio, the long-term behaviour of light curve may bear the imprint of a close binary companion to the radio-loud AGN. Indeed, the gradually decreasing trend is consistent with the expected spin--orbit precession which slowly increases the viewing angle of the jet.

Recent developments in extragalactic neutrino astronomy suggest that AGN with jets inclined close to our line of sight might be strong sources of the high-energy neutrinos reconstructed in the IceCube Neutrino Detector. Based on the properties of its VLBI jet, the binary parameters proposed by \citet{hu20}, and the merger-induced neutrino emission scenario proposed by \citet{Kun2017,Kun2019}, we found that Spikey could become an efficient high-energy neutrino source if the horizon of the secondary SMBH is rapidly rotating.

While the observed VLBI jet structure and the long-term trend in the flux density monitoring could possibly be explained with other effects, the consistency of both types of measurements with the Spikey binary parameters is remarkable, and can be considered as a support for the model of \citet{hu20}. The jet parameters could be determined with higher confidence and our values confirmed in the future with further frequent sensitive multi-epoch VLBI imaging observations. Our jet structural model involving eccentric orbit can later be applied for similar binary candidate AGNs with a jetted companion. 

\section*{Data availability}

The datasets underlying this article were derived from sources in the public domain as given in the respective footnotes.

\section*{Acknowledgements}

We thank Daniel D'Orazio for his comments on the manuscript and updates on the Spikey model parameters prior to publication. E.K. thanks the Hungarian Academy of Sciences for its Premium Postdoctoral Scholarship. K.\'{E}.G. was supported by the J\'{a}nos Bolyai Research Scholarship of the Hungarian Academy of Sciences and by the \'UNKP-19-4 New National Excellence Program of the Ministry of For Innovation and Technology. The National Radio Astronomy Observatory is a facility of the National Science Foundation operated under cooperative agreement by Associated Universities, Inc. 
We acknowledge the use of archival calibrated VLBI data from the Astrogeo Center database maintained by Leonid Petrov.
This research has made use of data from the OVRO 40-m monitoring program \citep{richards11} which is supported in part by NASA grants NNX08AW31G, NNX11A043G, and NNX14AQ89G, and NSF grants AST-0808050 and AST-1109911.
This research has made use of the NASA/IPAC Extragalactic Database (NED), which is funded by the National Aeronautics and Space Administration and operated by the California Institute of Technology.

\end{document}